\documentclass[letterpaper]{article}

\usepackage[T1]{fontenc}

\usepackage{geometry}
\usepackage{setspace}

\usepackage[style = chem-acs]{biblatex}
\usepackage{graphicx}
\usepackage{float}
\newfloat{scheme}{htbp}{los}
\floatname{scheme}{Scheme}
\floatname{chart}{Chart}
\newfloat{graph}{htbp}{loh}

\usepackage{chemformula} 
\usepackage[version = 4]{mhchem} 

\title{Non-equilibrium state during proton-deuteron exchange at a liquid-liquid interface}

\author{
	Tillmann~Buttersack$^{1\ast\dagger}$,
	Niclas~Sven~Mueller$^{1, 2 \dagger}$,
	Giulia~Carini$^{1}$,
    Henrik~Haak$^{1}$,\\
    Hanna~Bordyuh$^{3}$,
    Dipali~Singh$^{3}$,
    Sandy~Gewinner$^{1}$,
    Marco~De~Pas$^{1}$,\\
    Wieland~Schöllkopf$^{1}$,
    Martin~Wolf$^{1}$,
    Hendrik~Bluhm$^{1}$,
    Nils~Huse$^{3}$,\\
    Bernd~Winter$^{1}$,
    Gerard~Meijer$^{1}$,
    Alexander~Paarmann$^{1\ast}$\and
	\small$^{1}$Fritz Haber Institute of the Max Planck Society, Faradayweg 4--6, 14195 Berlin, Germany.\and
    \small$^{2}$Freie Universität Berlin, Department of Physics, 14195 Berlin, Germany.\and
	\small$^{3}$University of Hamburg, Luruper Chaussee 149, 22761 Hamburg, Germany.\and
	\small$^\ast$Corresponding authors. Email: buttersack@fhi.mpg.de or alexander.paarmann@fhi.mpg.de\and
	\small$^\dagger$These authors contributed equally to this work.
}

\begin{document}

\maketitle

\begin{abstract} \bfseries \boldmath
Proton-deuteron exchange is a very fast process, even across macroscopic length scales. Here we directly and quantitatively measure the formation of HDO within the first 100~microseconds of the reaction at the liquid–liquid interface between D$_2$O and H$_2$O, using a fast-flowing liquid flat jet combined with infrared spectroscopic imaging. We demonstrate that, at early stages HDO formation is reaction-limited, set by the low concentration of the hydroxide and hydronium ions that mediate the exchange. As the ion concentration rises, the rate rapidly approaches the diffusion limit. The reaction rate constant we extract is consistent with the picosecond timescale of the elementary proton–deuteron exchange. Access to these microsecond kinetics reveals a non-equilibrium state in the early H$_2$O/D$_2$O interface: the two liquids are fully mixed by diffusion, yet the HDO concentration remains well below equilibrium. Quantitative imaging of reactant and product concentrations at well-defined liquid–liquid interfaces, as introduced here, will enable the study of fast kinetics across a wide range of chemical reactions.
\end{abstract}

\noindent
 
\section{Introduction}

The most fundamental acid-base reaction is proton transfer between two water molecules by autoionization,\cite{Eigen1955} an extremely rare event that gives rise to the low concentrations of hydronium and hydroxide ions in pure water.\cite{Geissler2001} Once formed, these ions diffuse rapidly through consecutive proton transfers via the Grotthuss mechanism.\cite{Grotthus1806} The resulting balance between low ion concentration and fast ion diffusion is central to processes throughout biology and industry. Proton transfer and exchange therefore remain the subject of intense research, probed by infrared spectroscopy,\cite{Fournier2018, Fecko2003, Sarrazin2008, Nicodemus2011, Roberts2011, Cowan2005, Ji2010, Dahms2017, Thaemer2015, Wolke2016, Yuan2019} X-ray absorption spectroscopy,\cite{Ekimova2022} electron scattering,\cite{Yang2021} NMR spectroscopy,\cite{Meiboom1961} THz spectroscopy,\cite{Bruenig2022} and molecular dynamics simulations.\cite{Marx1999, Geissler2001, Hassanali2011, Chen2018, Yuan2019, Gomez2024} Most of previous work has addressed the microscopic details of the elementary proton transfer between two molecules.

Far less is known about proton transfer across interfaces, even though interfaces govern chemically specific charge and mass exchange throughout science and technology, from signal transduction in nervous systems and molecular transport in cell regulation to ion exchange in batteries and fuel cells and catalysis at solids.\cite{Crossley2010, Kaminski2017, Li2022} Interfaces between two liquids are a particularly important case. Interfaces between two immiscible liquids, such as water and oil, are comparatively straightforward to study,\cite{Piradashvili2016} whereas two inherently mixing liquids form an rapidly evolving interface, requiring continuously reformed steady state systems and fast probing techniques. A previous study of proton–deuteron exchange used a microfluidic system to monitor the exchange by Raman spectroscopy,\cite{Sarrazin2008} but with a time resolution on the order of seconds, leading to the conclusion that the exchange is governed solely by diffusion. Microfluidic flow rates, especially through thin transparent walls, are inherently limited by wall roughness and channel dimensions.\cite{Rands2006} Accessing the fast kinetics of a miscible liquid-liquid interface therefore requires a fundamentally different approach.

\begin{figure}[H]
    \centering
    \includegraphics[width=1\linewidth]{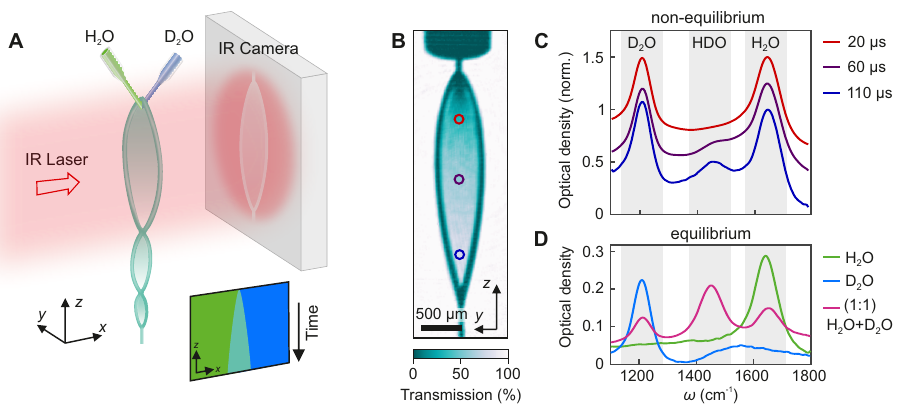}
    \caption{\textbf{Spatio-spectral imaging of proton-deuteron exchange at a liquid-liquid interface.} (\textbf{A}) Schematic of the experiment. Two impinging cylindrical microjets of H$_2$O and D$_2$O form a LS with a liquid-liquid interface. The inset schematically shows a cross-section of the H$_2$O (green) - D$_2$O (blue) LS as it evolves downstream, which leads to the gradual formation of HDO in the diffusively mixed region (turquoise) at the interface. The transmission of a narrowband and wavelength-tunable IR-FEL beam through the LS is imaged with an IR camera. (\textbf{B}) Transmission  image $T(y,z)$ for an H$_2$O/D$_2$O LS at the H-O-H bending vibration ($\omega_\mathrm{IR}=1640\,\mathrm{cm}^{-1}$). (\textbf{C}) Optical density $-\log_{10}(T)$ spectra extracted from three locations of the LS in B) marked by circles. The H-O-H and D-O-D bending vibrations of H$_2$O and D$_2$O, as well as the H-O-D bending vibration of HDO are well separated and allow for a quantitative analysis of the proton-deuteron exchange. (\textbf{D}) Reference spectra of pre-mixed solutions of H$_2$O (green), D$_2$O (blue), and a 50:50 mix of H$_2$O:D$_2$O (magenta).}
    \label{fig1}
\end{figure}

Here, we report the direct observation of proton-deuteron exchange across a water–water interface with millisecond time resolution (Fig.\,\ref{fig1}). We create the interface hydro-dynamically using a frictionless, fast-flowing ($\approx$ 22 m s$^{-1}$) flat jet, whose first thin leaf, the liquid sheet (LS, Fig.\,\ref{fig1}A), is formed by colliding two cylindrical microjets.\cite{Stemer2023, Buttersack2023} Since no turbulent mixing occurs, a well-defined interior liquid–liquid interface forms within the LS.\cite{Stemer2023, Buttersack2023} Schewe\,et\,al. previously imaged the chemiluminescence of an oxidation reaction at such an interface to obtain qualitative kinetics.\cite{Schewe2022} We probe the interface between H$_2$O and D$_2$O, where HDO forms rapidly (H$_2$O + D$_2$O $\rightleftharpoons$ 2 HDO) and reactants and products can be distinguished spectroscopically.\cite{Sarrazin2008, Nicodemus2011, Roberts2011, Dahms2017, Thaemer2015} Mixing equal amounts of H$_2$O and D$_2$O yields an equilibrium concentration of $c_{HDO,eq} \approx 27.5$\,M at 298\,K ($K_{eq} = 3.85$,\cite{Bigeleisen1973} i.\,e., a H$_2$O:D$_2$O:HDO ratio of 1:1:2). The two liquids mix efficiently, and an initially sharp interface softens quickly. Nevertheless, due to the steady-state within the LS and its high velocity, the travel distance from the impinging point (millimeters) translates into reaction and diffusion times of 10–-100\,$\mu$s,\cite{Schewe2022} beyond the reach of the stopped-flow method commonly used to study fast kinetics.\cite{Chan1997, Kise2014}

Specifically, we employ mid-infrared (1100–1800 cm$^{-1}$) spectroscopic imaging of the H$_2$O/D$_2$O interface within the LS, an approach highly selective to the H$_2$O, D$_2$O, and HDO bending vibrations. The absorption amplitudes give quantitative access to the evolution of reactants and products over the $\approx$100\,$\mu$s the molecules travel from the impingement point to the bottom of the LS. From these data we determine the HDO concentration along the interface, the effective diffusion coefficient, and the rate constant for HDO formation yielding an estimate of the average proton–deuteron exchange time. Moreover, we discuss the potential of this approach for other fast reactions at liquid–liquid interfaces.

\section{Results and Discussion}\label{sec2}
We employed a tunable, narrow-band infrared free-electron laser (IR-FEL) to perform mid-infrared spectroscopic imaging of water LSs as illustrated in Fig.~\ref{fig1} (see SI sect.~2). We used a wide-field imaging configuration, yielding transmission spectra at each point in the LS with a spatial resolution of 16\,$\mu$m. The reaction is initiated at the
impinging point of the cylindrical jets at the top of the LS. Directly below the impinging point, one side of the LS consists of pure D$_2$O, and the other of pure H$_2$O, with HDO gradually forming at the interface along the flow direction, see inset of Fig.~\ref{fig1}A. The down-stream distance to this point translates into the reaction time which can be calculated from the known flow velocity.\cite{Schewe2022} At the flow velocity of 22$\pm$4~m/s in our experiment, the spatial resolution of 16\,$\mu$m corresponds to an instrumental time resolution of 1.6\,$\mu$s. Notably, however, accounting for the finite vertical size of the impinging microjets of $\approx~64~\mu$m/$\cos(\theta)$, with $\theta = 48^{\circ}$ the impinging angle, results in an effective time resolution of $\approx~10~\mu$s.

A transmission image of an H$_2$O/D$_2$O LS taken at the H-O-H bending mode frequency $\omega_{\mathrm{IR}}$\,=\,1640\,cm$^{-1}$ is presented in Fig.~\ref{fig1}B. Using the known optical properties of H$_2$O,\cite{Hale1973} we can estimate the optical thickness of \ce{H2O} (Methods Fig.~\ref{fig:fig5}) which is strongly affected by the typical behavior of downstream thinning of the LS.\cite{Buttersack2023} The full chemical composition at any given position in the LS, on the other hand, can be monitored by scanning the IR-FEL wavelength, and thereby extracting the vibrational spectra at different positions in the LS. In Fig.~\ref{fig1}C we show three spectra at the respective positions marked by circles in Fig.~\ref{fig1}B, corresponding to early, intermediate, and late reaction times, respectively. Upstream in the LS (red curve in Fig.~\ref{fig1}C), we observe only two resonances of equal magnitude at the D$_2$O ($\omega_{\mathrm{IR}}=$1210~cm$^{-1}$) and H$_2$O ($\omega_{\mathrm{IR}}=$1640~cm$^{-1}$) bending mode frequencies, respectively. Further downstream (purple curve in Fig.~~\ref{fig1}C), a third resonance emerges at $\omega_{\mathrm{IR}}=$1450~cm$^{-1}$ which reports on the formation of HDO, becoming even more pronounced at the bottom of the sheet (blue curve). Reference spectra taken for LSs using the same liquid in both cylindrical jets, either pure H$_2$O, pure D$_2$O, and fully equilibrated 1:1:2 \ce{H2O}:\ce{D2O}:HDO, respectively, are shown in Fig.~\ref{fig1}D, confirming the assignment of the resonances in the H$_2$O/D$_2$O system.

\begin{figure}[H]
    \centering
    \includegraphics[width=1\linewidth]{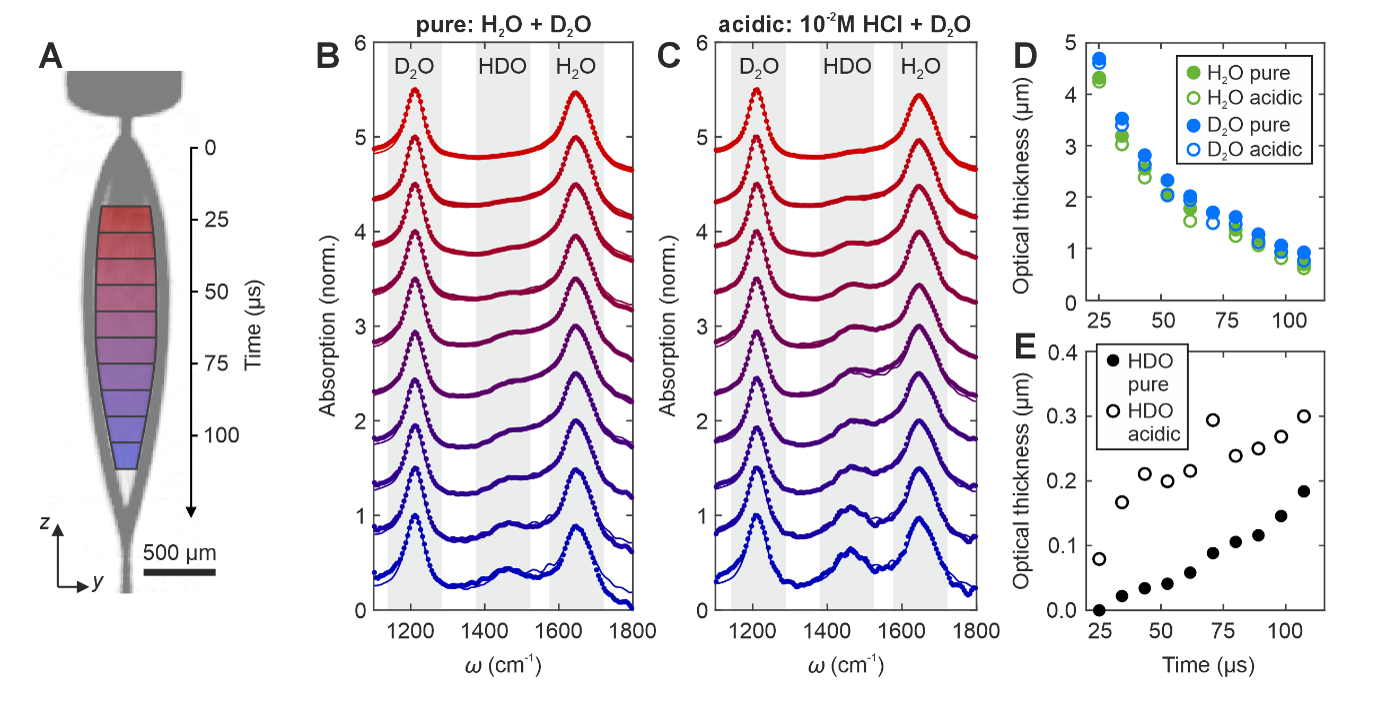}
    \caption{\textbf{Temporal evolution of the infrared absorption spectra of the H$_2$O/D$_2$O interface.}
    (\textbf{A}) Optical density image of H$_2$O/D$_2$O LS at $\omega_\mathrm{IR}=1640\,\mathrm{cm}^{-1}$ with binned areas (colored boxes) that are used to extract IR spectra from averaging over all pixels in each area corresponding to different times of the reaction, with 10 $\mu$s time resolution.
    (\textbf{B}) Average IR spectra from binned areas in A) recorded at the H$_2$O/D$_2$O interface. Besides the vibrational modes of H$_2$O (1640 cm$^{-1}$) and D$_2$O (1210 cm$^{-1}$) the band of HDO (1450 cm$^{-1}$) is evolving with reaction time. The lines show fits to the measured data points (symbols) using a linear combination of reference spectra (Methods), where the colors refer to the bins in A).    
    (\textbf{C}) Same as B) but for a 10$^{-2}$\,M\,HCl/D$_2$O LS, which leads to an accelerated formation of HDO.
    (\textbf{D}) Optical thickness of H$_2$O (green) and D$_2$O (blue) and (\textbf{E}) of HDO as a function of time. Full symbols refer to the H$_2$O/D$_2$O LS in B), and empty symbols to the HCl/D$_2$O interface in C).
    }
    \label{fig2}
\end{figure}

To quantitatively analyze the reaction dynamics, we extracted transmission spectra as a function of reaction time, see Fig.~\ref{fig2}. Here, we excluded the data for the first $\approx$20~$\mu$s due the low transmission in this thickest region of the LS. With the spatial binning as shown in Fig.~\ref{fig2}A, this approach provides time-resolved spectra over a range of $>$100~$\mu$s with a resolution of 10~$\mu$s, as shown in Fig.~\ref{fig2}B.  These data show a clear evolution of the HDO spectral signatures with reaction time, as expected from Fig.~\ref{fig1}C. A comparison of the spectra taken at the bottom of the LS shows that they are considerably different from those of fully equilibrated \ce{H2O}/\ce{D2O} (compare Fig.~\ref{fig2}B with Fig.~\ref{fig1}D). This observation indicates that the reaction has not yet reached equilibrium within the 100~$\mu$s observation time of our experiment for pure H$_2$O/D$_2$O. However, the reaction accelerates when adding reactive species (H$^+$, OH$^-$) to one or both of the impinging microjets. The results are shown in Fig.~\ref{fig2}C where we replaced pure H$_2$O by an aqueous solution of 10$^{-2}$\,M HCl. In these data, the HDO signatures are overall more pronounced than for pure water, indicative of faster reaction kinetics. 

The amount of HDO produced at the interface is quantified by extracting the optical thickness of each species as a function of time by fitting a linear combination of reference spectra of H$_2$O, D$_2$O, and HDO to each spectrum in Fig.~\ref{fig2}B,C, see Fig.~\ref{fig:FigS2}. The resulting (optical) thickness for each of the species is plotted in Fig.~\ref{fig2}D,E. For both scenarios, neutral and acidic-water LSs, the optical thickness of H$_2$O and D$_2$O rapidly decays due to the overall thinning of the LS along the -$z$-direction, as well as due to chemical conversion to HDO.\cite{Buttersack2023} At the same time, the optical thickness of HDO steadily increases as the reaction proceeds (Fig.~\ref{fig2}E). For the HCl aqueous solution (open symbols), the HDO optical thickness grows more quickly initially  compared to the pure water case (closed symbols).

\begin{figure}
    \centering
    \includegraphics[width=\columnwidth]{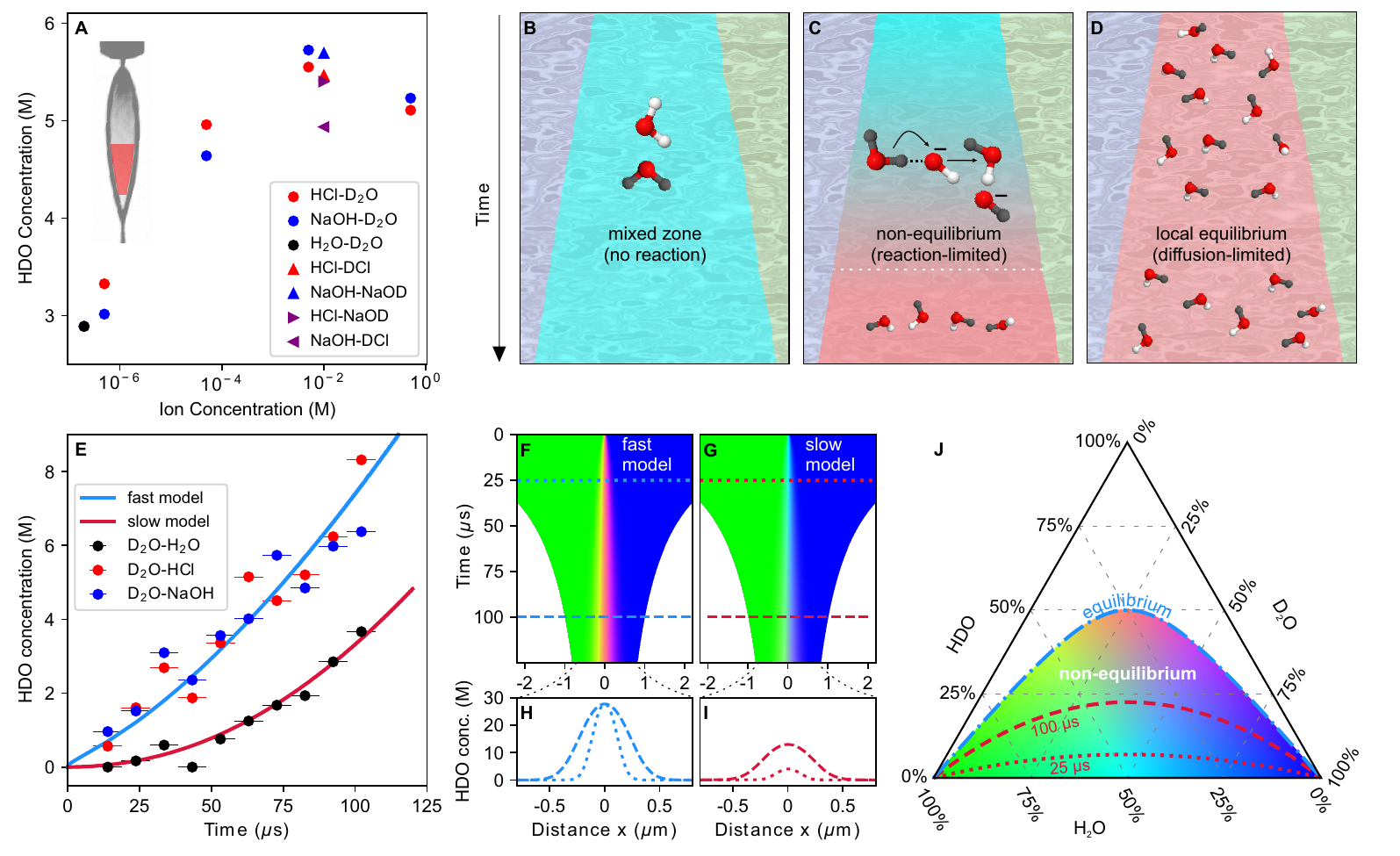}
    \caption{\textbf{Non-equilibrium state at the liquid-liquid interface.}    
    (\textbf{A}) Average concentration of HDO in the bottom half of the LS (see inset) as function of available hydroxide and hydronium ions, using D$_2$O/(acidic/basic H$_2$O) (dots) and (acidic/basic D$_2$O)/(acidic/basic H$_2$O) (triangles) combinations, see legend.    
    (\textbf{B}--\textbf{D}) Conceptual illustration of the three scenarios of the evolution of the \ce{H2O}/\ce{D2O} interface:
        (\textbf{B}) (hypothetical) no HDO formation, mixing only,
        (\textbf{C}) intermediate rate of HDO formation leading to a non-equilibrium state, and
        (\textbf{D}) very fast HDO formation resulting in a local equilibrium.
    (\textbf{E}) Concentration of HDO as a function of reaction time. The solid lines represent the simulated dynamics, based on integrating (along $x$) the concentration distributions as shown in F,G). The limited availability of ions causes severe slowing down of the dynamics for the pure H$_2$O/D$_2$O system (crimson curve), compared to fast model with higher ion concentration (light blue curve). 
    (\textbf{F,G}) Simulated distribution of H$_2$O, D$_2$O, and HDO as a function of time and distance from the liquid-liquid interface, with parameters fitted to reproduce the experimental data in E). (\textbf{F}) Fast model: Assuming near-instantaneous HDO formation results in a local equilibrium concentration at each position of the evolving interface, cf. sub-panel D). (\textbf{G}) Slow model: In the case of H$_2$O and D$_2$O, the local equilibrium concentration of HDO is not yet reached, cf. sub-panel C).
    (\textbf{H,I}) HDO concentrations at an early (dotted line) and late (dashed line) time. The fast model (\textbf{H}) results in local equilibrium concentration of HDO at any stage where 27.5\,M corresponds to 1:1:2 \ce{H2O}:\ce{D2O}:HDO. For the slow model (\textbf{I}), the HDO concentrations stay below the equilibrium values. (\textbf{J}) Ternary color-scale for F,G) of the non-equilibrium part of the \ce{H2O}/\ce{D2O} system, where the light-blue dashed-dotted line marks the HDO equilibrium concentrations as observed for the fast model in H), while red lines indicate the mixture after 25 and 100 $\mu$s as predicted by the slow model.
    }
    \label{fig3}
\end{figure}

Our data suggest that HDO production across the liquid-liquid interface is not always limited simply by molecular diffusion, but instead can enter a reaction-limited regime at early reaction times and at low concentrations of hydroxide and hydronium ions acting as reaction intermediates for the proton-deuteron exchange, see SI for details. To confirm this hypothesis, we acquired data for a series of acidic and basic ion concentrations of heavy and light water, see Fig.~\ref{fig3}A. We show the average HDO concentration in the second half of the LS which clearly increases with increasing ion concentration. A combination of acid and base solutions (purple triangles) leads to neutralization at the liquid-liquid interface, which likely explains the slightly smaller HDO production rate in those cases compared to purely acidic or basic systems (blue and red symbols). Overall, however, our data clearly show that the rate of HDO production across the interface is determined by the ion concentration, and is at the same time largely independent of which ions (OH$^{-}$, OD$^{-}$, H$_3$O$^+$, D$_3$O$^+$) are provided at increased concentrations.

Based on these observations, we employ a simple kinetic model to describe the experimental results. We assume that HDO formation at the \ce{H2O}/\ce{D2O} interface is governed by two processes: molecular water diffusion across the interface, and HDO formation kinetics within the diffusively mixed volume. In Fig.~\ref{fig3}B--D, we conceptually illustrate three different scenarios for the evolution of the \ce{H2O}/\ce{D2O} interface based on this model. In Fig.~\ref{fig3}B, hypothetically, \ce{H2O} (light green) and \ce{D2O} (light blue) diffuse into each other without forming HDO (no reaction). This would result in a mixed interface region (cyan) diffusively growing over time. Conversely Fig.~\ref{fig3}D illustrates the other limiting case, where the local equilibrium concentration of HDO is reached instantaneously in the mixed region (red) for very fast HDO formation (diffusion-limited). In an intermediate case, Fig.~\ref{fig3}C, the proton-deuteron exchange occurs on similar time scales as diffusion. In that case, the local equilibrium is not reached instantly, resulting in a reaction-limited steady-state of the interface, constituting a non-equilibrium state of the proton-deuteron exchange reaction. 

The formation of HDO molecules requires a proton or deuteron transfer between a water molecule and hydroxide or hydronium ions (Fig.\,3C), see SI for all possible direct reaction pathways. These ions are present in water due to the autoionization. Importantly, in order to reach the equilibrium concentration of HDO, which is 50\% HDO in a 1:1 mixture of H$_2$O and D$_2$O, many of these proton-deuteron exchange reaction steps are needed, and the overall reaction accelerates with higher concentration of hydronium and hydroxide ions in the mixed interface region. As an example, the reaction between \ce{D2O} and OH$^-$ is depicted in Fig.~\ref{fig3}C. This reaction (and their counterparts with \ce{H2O}, or hydronium ions, respectively) are expected to locally follow first-order kinetics, since the concentrations of \ce{D2O} and \ce{H2O} limit the reaction rate when approaching equilibrium. Once local equilibrium is reached in the mixed zone, further formation of HDO is limited by diffusive expansion of the mixed region. 

To quantitatively describe the experimental data, we devise a simple reaction-diffusion model.\cite{Hundsdorfer2003} We assume that molecular water diffusion with a single effective diffusion constant $D_{\mathrm{eff}}$ (assumed the same \ce{H20} and \ce{D2O} here) controls the molecular mixing at the interface, forming an expanding mixed layer with a local ratio of \ce{H2O} and \ce{D2O}. The reaction that leads to the formation of HDO at each location is described as:
\begin{equation}
    \ce{H_2O + D_2O <=>[$k_f$][$k_b$] 2 HDO}
    \label{eq:reaction}
\end{equation}
where the forward and backward rate constants $k_f$ and $k_b$ respectively are expected to depend on the ion concentration. Their ratio, however, is determined by the equilibrium constant $K_{eq}$ such that $k \equiv  k_f = K_{eq}\cdot k_b$. Under these assumptions, our model only has two free parameters $D_{\mathrm{eff}}$ and $k$, and we can numerically solve the corresponding rate equation. Integrating the resulting spatial distributions of \ce{H2O}, \ce{D2O}, and HDO across the time-dependent LS thickness (see Methods) provides the total HDO concentration as a function of time. We use this model to fit our experimental data as shown in Fig.~\ref{fig3}E for a \emph{fast model} (light blue) and a \emph{slow model} (crimson), reproducing very well the dynamics observed in experiment for the acidic or basic and the pure water LS, respectively. 

Fitting the fast model, we assumed that the local equilibrium is reached immediately in each simulation time step, corresponding to the reaction being much faster than diffusion. This model can be employed for sufficiently high concentrations $\geq$10\,mM of hydroxide or hydronium ions, as evidenced by the saturation of the reaction speed with ion concentration observed in our data (Fig.~\ref{fig3}A and Fig.~\ref{fig:16plots}). This leaves the effective diffusion constant $D_{\mathrm{eff}}$ as the only free parameter for the fast model. Fitting this simulation to our experimental data with 10$^{-2}$\,M HCl or NaOH solutions, we extract an effective diffusion coefficient $D_{\mathrm{eff}}=(2.8\pm0.9)\cdot10^{-10}$\,m$^2$s$^{-1}$. Interestingly, this value is lower than the self-diffusion coefficient of bulk water ($\approx 2\cdot10^{-9}$\,m$^2$s$^{-1}$),\cite{Mills1973} which likely is related to the details of three-dimensional mass transport within the fast-flowing LS and into the rims of the leaf,\cite{Choo2007} which are not included in our model. Assuming that the pure \ce{H2O}/\ce{D2O} LS experiences the same effective diffusion, we fit the slow model with a finite reaction rate $k$ to the pure water data, crimson curve and black dots in Fig.~\ref{fig3}E, respectively, which  yields excellent agreement for an HDO formation rate constant of $k = (9.5\pm3.2)\cdot10^3$\,s$^{-1}$.

In Figs.~\ref{fig3}F,G, we compare the simulated evolution of the interface for the two different models shown in Fig.~\ref{fig3}E. We plot the concentration profiles of H$_2$O, D$_2$O, and HDO as a function of flow time from the impinging point, and distance $x$ from the interface, where we also account for the downstream thinning of the LS. We employ a triangular color bar (Fig.~\ref{fig3}J) to track the concentrations of all three species. For the fast model (Fig.~\ref{fig3}F), the local HDO equilibrium is always reached, while in the slow model (Fig.~\ref{fig3}G), the system is not yet in equilibrium. This can be seen clearly in Fig.~\ref{fig3}H,I where we plot cuts through the distribution at early and late times, dotted and dashed line in Figs.~\ref{fig3}F,G, respectively. For the fast model and within the time resolution of our experiment, the HDO concentration is always maximal at the interface where we expect a 1:1 mixture of H$_2$O and D$_2$O, and the HDO distribution gradually broadens due to the diffusive mixing of H$_2$O and D$_2$O. For the slow model, the same diffusive mixing takes place, yet the HDO concentration is well below the local equilibrium value, and only gradually increases downstream. This behavior can also be understood in the triangular color bar (Fig.~\ref{fig3}J): the fast model results in local equilibrium concentrations of HDO that depend on the local mixing ratio of H$_2$O and D$_2$O, taking values only along the light blue dash-dotted line. For the slow model, however, we also find below-equilibrium HDO concentrations, exemplified by the dotted and dashed crimson lines at early and late times, respectively. The reaction-limited (non-equilibrium) state approaches diffusion-limited kinetics after a few ms, see SI Fig.~\ref{fig:short_vs_long}.

Clearly, our model reproduces the experimental data very well despite neglecting any ionic diffusion across the interface which is expected to vary significantly for each ion species (differently deuterated hydronium and hydroxide species) and their respective initial concentrations on either side of the interface.\cite{weingartner1990anomalous} In principle, any hydronium or hydroxide ion diffusing across the interface would lead to rapid formation of an HDO molecule (see SI Sect.~\ref{sec:reactions}). However, each ion which crossed over to the other side of the interface before molecular mixing, such as OD$^-$ into H$_2$O,  can only mediate formation of a single HDO molecule. Considering the rather low ion concentrations employed in our experiments, such ion diffusion via the Grotthuss mechanism across the interface and subsequent HDO reaction is insufficient to explain the high HDO concentrations observed. In contrast within our model, conversion only happens in the layer mixed by molecular water diffusion, where each ion can facilitate conversion of many water molecules into HDO. These considerations are consistent with the possibly surprising observation of the HDO formation rates being largely independent of the ion species and only depending on their concentration (Fig.~\ref{fig3}A) in our experiments.

Notably, we can use the HDO formation rate $k$ extracted from our data to estimate the time scale $t_x$  of an individual proton-deuteron exchange process. In the diffusively mixed layer, Eq.~\ref{eq:reaction} results in pseudo-first-order kinetics for HDO formation:\cite{Canet2016, Grimaldi2026}
\begin{equation}
    f_{eq.}(t)  = \frac{c_{HDO}(t)}{c_{HDO}(eq.)}= 1 - e^{-k \cdot t},
    \label{eq:thop}
\end{equation}
where $f_{eq.}(t)$ is the ratio of the local HDO concentration and the local equilibrium concentration, the latter in turn depending on the local ratio of \ce{H2O} and \ce{D2O}. The time scale of the individual proton-deuteron exchange process can be estimated by solving Eq.~\ref{eq:thop} for $t_x$ such that $c_{HDO}(t_x) = c_{ion} = 2 \cdot 10^{-7}$\,M from auto-ionization, corresponding to one HDO formed by each hydroxide and hydronium ion. We extract an average time for the individual proton-deuteron exchange process of $t_x \approx 0.6 - 1.1$\,ps. This average exchange time is consistent with the proton hopping time extracted from ultra-fast experiments and MD-simulations;\cite{Meiboom1961, Yuan2019, Gomez2024}  and is also very similar to hydrogen-bond lifetimes.\cite{Luzar1996, Yang2021}

The difference between the \ce{D2O}/\ce{H2O} and the pH-adjusted (HCl, NaOH) systems is significant in our experiment. The pH-adjusted system allows us to determine the effective diffusion across the fast flowing LS, while the \ce{D2O}/\ce{H2O} system provides access to the picosecond timescale of the proton-deuteron exchange. This is remarkable given that the temporal resolution of our experiment is on the order of microseconds. The fast reaction-limited kinetics are accessible because the measured reaction rate constant is scaled down by the large ratio ($\approx 10^6$) between the equilibrium concentration of HDO in the 1:1 mixture and the very low concentration of hydroxide and hydronium ions in pure water.

In our experiment, we directly measure the absolute amount of product, rather than monitoring the process at the molecular level. This means that the kinetics of a very fast reaction become accessible because fast-flowing LSs exploit the scaling of the reaction rate with the reactant concentrations. More generally, to measure kinetics at a liquid-liquid interface, the reaction rate constant $k$ must be compatible with the spatial resolution of the imaging system and the flow velocity of the LS ($k\approx10^3-10^4\,\textrm{s}^{-1}$). As we show here, even fast reactions such as proton exchange can be brought into this range by making use of scaled kinetics governed by a pre-equilibrium (e.g., by autoionization). Therefore, we foresee a large potential for this approach to study fast kinetics at liquid-liquid interfaces, such as for instance the proton exchange between fatty acids and aqueous systems at liquid-liquid interfaces, as prototype systems for mimicking cell membranes. The proton-transfer dynamics accessed here also have direct implications for organic synthesis in biphasic systems,\cite{Dedovets2021} as well as reactions where proton transfer is rate-limiting, including amide synthesis,\cite{pattabiraman2011rethinking} aldol condensation,\cite{trost2010direct} and phase transfer catalysis.\cite{ooi2007recent}

\section{Conclusion}\label{sec13}

We used mid-infrared spectroscopic imaging to investigate one of the most fundamental chemical reactions in nature: the formation of semiheavy water from H$_2$O and D$_2$O. Employing a fast-flowing liquid sheet to form a well-defined liquid-liquid interface, we followed the formation of HDO within the first 100\,$\mu$s at the interface between D$_2$O and H$_2$O.
Although proton-deuteron exchange is usually considered a fast, diffusion-limited reaction, we show that at the H$_2$O/D$_2$O interface it is instead reaction-limited by the scarcity of hydroxide and hydronium ions in pure water. This gives access to a spatially resolved non-equilibrium state of the interface, with HDO concentrations well below equilibrium. It is the microsecond time resolution of our experiment that lets us capture this non-equilibrium state, while we expect diffusion-limited kinetics to dominate after a few milliseconds. At higher ion concentrations (e.g., 10\,mM), the reaction accelerates and, within our experimental resolution, HDO formation becomes diffusion-limiting.  From the interplay of ion and product concentration, we connected the finite rate of HDO formation to elementary proton-deuteron exchange processes that occur on the picosecond timescale, despite the time resolution of the experimental setup being on the microsecond scale. The approach introduced here opens up new opportunities for studying fast kinetics across a wide range of liquid-liquid interface systems.

\renewcommand{\thefigure}{M\arabic{figure}}
\setcounter{figure}{0}

\section*{Materials and Methods}

\subsection*{Sample preparation}
Deuteriumoxide (D$_2$O) with a purity of 99.9\% as well as 1\,M solutions of HCl and NaOH were purchased from VWR. Concentrated solutions of DCl and NaOD in D$_2$O were obtained from Sigma-Aldrich (35\% and 40\%, respectively). The acid and base solutions with concentrations used in the experiments were prepared by dilution with H$_2$O (Ultrapure, 18 M$\Omega \cdot$ cm) and D$_2$O, respectively. The solutions were pumped with a two channel HPLC pump (Shimadzu\,LC-40B\,XR with degasing unit) into the flatjet system. Details of the flatjet system have been described elsewhere.\cite{Stemer2023, Buttersack2023} The flowrate of each sample line was 2.8\,mL/min, respectively. The cylindrical jets had a diameter of 64\,$\mu$m, therefore the velocity of each cylindrical jet was $u_{LJ}$:  15\,m s$^{-1}$. The two cylindrical jets were impinged with an angle of 48\textdegree. The velocity in the LS is about 1.5 times as large as the velocity of the cylindrical microjets.\cite{Choo2007} Therefore, the velocity in the LS $u_{LS}$ is estimated as 22$\pm$4\,m$\cdot$\,s$^{-1}$.

\subsection*{Infrared spectral imaging}
\begin{figure}[H]
    \centering
    \includegraphics[width=1\linewidth]{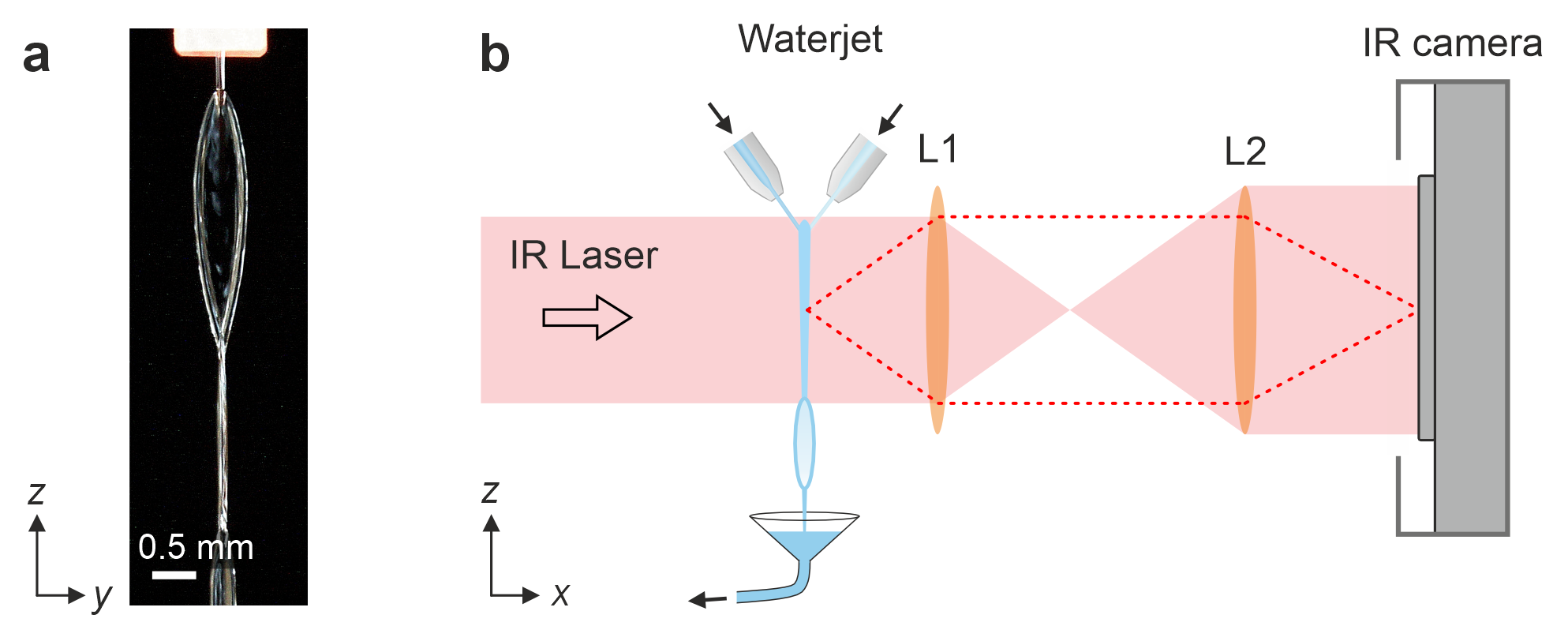}
    \caption{Experimental setup for spectroscopic wide-field infrared imaging. (a) Photograph of a typical flat jet with the imaging plane parallel to its first leaf. (b) Wide-field imaging geometry using two ZnSe lenses L1 and L2. In the experiment, we used a 5x magnification (not shown here for clarity) by using focal lengths of $f_1=5\,$cm and $f_2=25\,$cm.}
    \label{fig-setup}
\end{figure}

Infrared spectral imaging was conducted with the IR free-electron  laser (IR-FEL) of the Fritz Haber Institute Berlin, that provides tunable ($\omega_{IR}$ = 200 – 3000 cm\,$^{-1}$) and narrowband ($\approx$0.6\% FWHM of $\omega_{IR}$) infrared radiation. The laser has a micro-/macro-pulse structure, with micro pulses: 1\,GHz repetition rate, $\approx$5\,ps pulse length, $\approx$5\,$\mu$J pulse energy; macro pulses: 10\,Hz repetition rate, 8\,$\mu$s pulse length, $\approx$80\,mJ pulse energy in our experiments. The FEL laser beam was collimated to a beam diameter of $\approx$5\,mm and guided in transmission through the water jet (Fig.\,\ref{fig-setup}). The transmitted light was then imaged onto an IR camera with two ZnSe lenses that magnified the image x5. We used a pyroelectric array beam profiler (Pyrocam IV, Ophir Optronics Solutions) as IR camera with 320x320 pixels and 25.6\,mm x 25.6\,mm active area, corresponding to an 80\,$\mu$m effective pixel size. The camera was triggered to record images only when FEL macropulses arrived with an integration time that included the entire macro pulse. 
For spectral imaging, the IR frequency was scanned from $\omega_{IR}$ = 1100\ cm$^{-1}$ to 1800\,cm$^{-1}$ in steps of 5\,cm$^{-1}$. At each IR frequency we recorded 10 images of subsequent macro pulses that were averaged. After or before each frequency sweep, we recorded reference images with the same settings by moving the water jet out of the IR laser beam. By dividing the images with and without the jet, we obtained transmission images, as shown, e.g., in Fig.\,1b. Each frequency sweep took about 15\,min. Between each measurement series with acids or bases we repeated measurements on the pure \ce{H2O}:\ce{D2O} waterjet to ensure that all water tubes were clean and the previous reaction rate was reproduced. The temperature in the experimental hall was 22-23\textcelsius.

\subsection*{Analysis of infrared images and spectra}
The images and spectra were analyzed with home-written code in Matlab. Each transmission image was corrected by vertical line-wise scaling, such that the transmission beside the water jet was set to 100\%. For each image, we used an algorithm to extract the interior of the LS (Fig.\,2A) and spectra were extracted for each image pixel from the frequency sweeps. The optical density OD = -log$_{10}$\textit{T} was then calculated for each image pixel. The OD spectra were averaged in selected regions of the LS as described in the main text. 
To reduce noise, the OD spectra were further smoothed using a moving average filter while keeping the increase in spectral line width below 10\%. Reference spectra of \ce{H2O}:\ce{H2O}, \ce{D2O}:\ce{D2O}, as well as a jet that consisted of a 50:50 mixture of \ce{H2O}:\ce{D2O} were extracted from the entire interior of the LS for Fig.\,1d, or for each bin shown in Fig.\,2a for the analysis underlying Fig.\,2\,and\,3. As the \ce{H2O}:\ce{D2O} mixture consists to 50\% of HDO and 25\% \ce{H2O} and \ce{D2O} respectively, a reference spectrum of pure HDO was estimated as $2\cdot$[\ce{H2O}:\ce{D2O} – (\ce{H2O}:\ce{H2O} + \ce{D2O}:\ce{D2O})/4]. This reference spectrum is called HDO in the following. The spectra of the \ce{H2O}:\ce{D2O} flat jets with different ion concentrations were fitted with a linear combination of the reference spectra and a constant background (Fig.\,\ref{fig:FigS2}). In the fits, the \ce{H2O}:\ce{H2O} reference spectrum was shifted by +7 cm$^{-1}$ and the HDO spectrum by +5 cm$^{-1}$ to obtain the best spectral match to the \ce{H2O}:\ce{D2O} spectra. To improve the fitting accuracy in the spectral range of the weak HDO peak, the fits were weighted by a Gaussian centered at 1450 cm$^{-1}$, see gray shaded area in Fig.~\ref{fig:FigS2}. The spectra for 1M\,HCl:\ce{D2O} were fitted by using HCl:HCl as reference spectrum instead of \ce{H2O}:\ce{H2O}. From these fits, we obtained the relative concentration of \ce{H2O}, \ce{D2O} and HDO at different positions in the flat jet.

\begin{figure}
    \centering
    \includegraphics[width=1\linewidth]{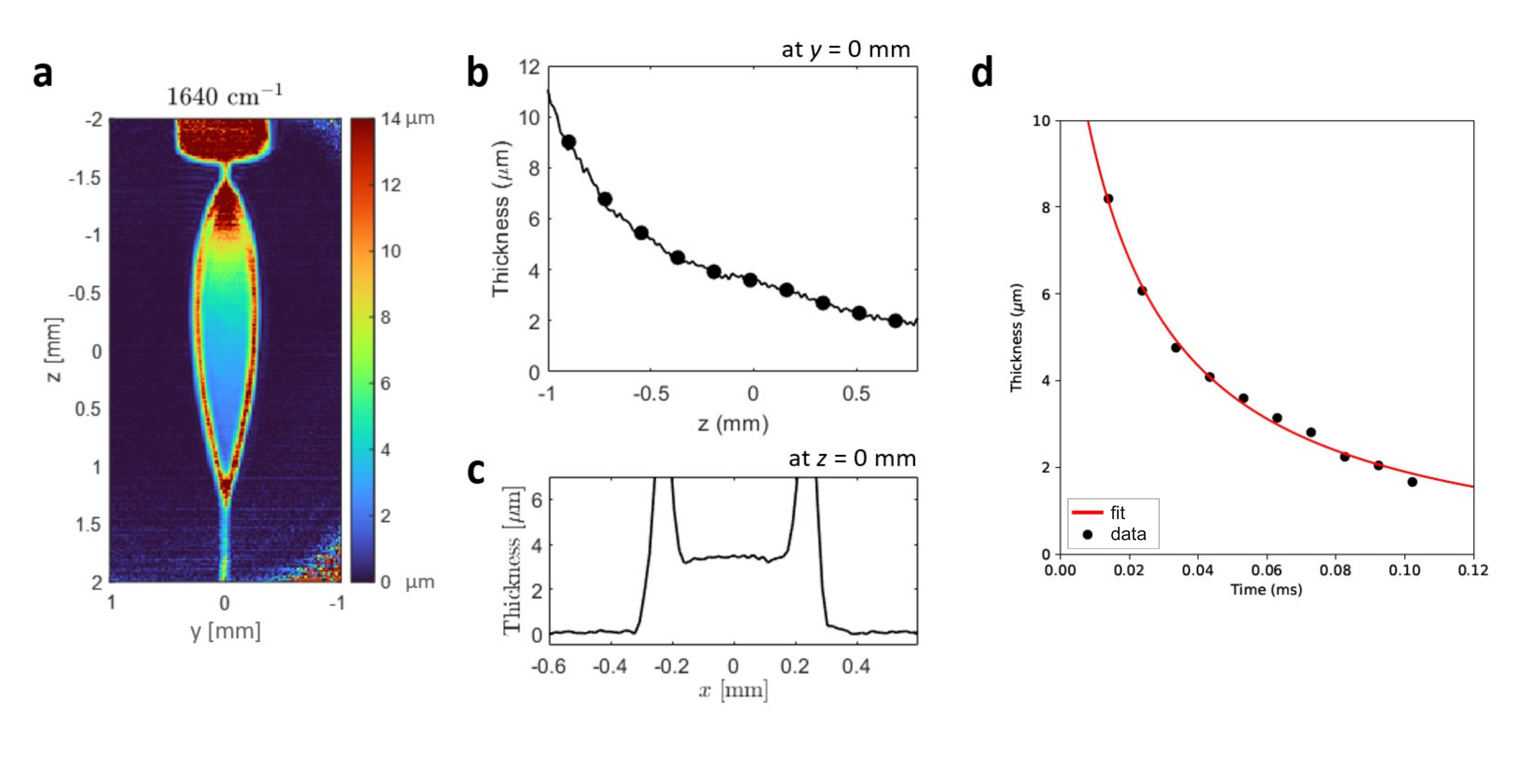}
    \caption{Thickness of LS in (a) the entire LS, (b) at central axis of the LS along the direction of flow and (c) in center of LS perpendicular to the direction of flow. (d) Fit of the experimental thickness (black dots) as function of time with Hasson-Peck model (red line).}
    \label{fig:fig5}
\end{figure}

We determined the optical thickness \textit{d} of the \ce{H2O}:\ce{H2O} jet in Fig.\,1b, using 
\cite{Tomlin1968}
\begin{equation}
    d = -\mathrm{ln} \left( \frac{T}{16} \frac{T [(1+n)^2 + k^2]^2}{n^2+k^2} \right) \frac{c}{2k\omega},
\end{equation}
with $n+ik$ the complex refractive index of water. We determined the thickness from images at $\omega_{IR}$ = 1600, 1640 and 1700 cm$^{-1}$ using the refractive index of water at 25\textdegree C  from Ref. \cite{Hale1973} and subsequent averaging, Fig.~\ref{fig:fig5}. 
The optical thicknesses in Figs.~\ref{fig2}D,E were obtained by multiplying the thickness of the H2O:H2O liquid jet in Fig.~\ref{fig:fig5} with the fractions of \ce{H2O}, \ce{D2O}, and HDO from fits of the spectra in Figs.~\ref{fig2}B and C, assuming that the thicknesses of all liquid jets are the same. We use the term “optical thickness”, as these thicknesses are estimated from optical measurements. The HDO concentrations reported in Figs.~\ref{fig3}A,E were instead obtained by multiplying \ce{H2O}, \ce{D2O}, and HDO fractions (from fits) with the concentration 55.5M of water. These values are thus independent of the thickness measurement in Fig.~\ref{fig:fig5}.

The thickness of the LS at the central axis as a function of time $t$ (corresponding to position $z$) can be fitted with a Hasson-Peck model (red curve in Fig.\,\ref{fig:fig5}):\cite{Hasson1964}
\begin{equation}
    d(t) = a_1/(t+a_2)-a_3, 
    \label{eq:thick}
\end{equation}
with a $a_1 = 0.237\,\mathrm{\mu m}\cdot\mathrm{ms}^{-1}, a_2 = 0.016\,\mathrm{ms}$,  and $a_3 = 0.0525\,\mathrm{\mu m}$.

\subsection*{Simulation of the reaction-diffusion system}

We numerically solve the one-dimensional reaction-diffusion system\cite{Hundsdorfer2003} generally defined as:
\begin{equation}
    \frac{\partial \mathbf{f}(x,t)}{\partial t} = D \frac{\partial^2}{\partial x^2} \mathbf{f}(x,t) + \mathbf{R}(\mathbf{f}(x,t)),
\end{equation}
where specifically, $\mathbf{f}(x,t) = \{f_{H_2O}(x,t),f_{D_2O}(x,t),f_{HDO}(x,t)\}$ are the mole fractions of \ce{H2O}, \ce{D2O}, and HDO at location $x$ and time $t$, and $D$ is the diffusion constant. The reaction term $\mathbf{R}(\mathbf{f})$ is expressed as:
\begin{equation}
\mathbf{R}(\mathbf{f}) = 
\begin{pmatrix}
    -k(f_{H_2O} f_{D_2O} - \frac{1}{K_{eq}} f_{HDO}^2) \\
    -k(f_{H_2O} f_{D_2O} - \frac{1}{K_{eq}} f_{HDO}^2) \\
    2k(f_{H_2O} f_{D_2O} - \frac{1}{K_{eq}} f_{HDO}^2)
\end{pmatrix}
\end{equation}
where we omitted the $(x,t)$-dependence for clarity. $K_{eq} = 3.85$ is the equilibrium constant of HDO,\cite{Bigeleisen1973} and $k$ is the reaction rate. Initial conditions are set such the $f_{H_2O}(x<0,t=0) = 1$,~$f_{H_2O}(x>0,t=0) = 0$,~$f_{D_2O}(x<0,t=0) = 0$,~$f_{D_2O}(x>0,t=0) = 1$, and $f_{HDO}(x,t=0) = 0$. We complementarily use the Matlab differential equation solver $pbede$ as  well as an equivalent Python (BDF solver) implementation to numerically solve the problem for parameters $D$ and $k$, both yielding the same results.

To extract the concentrations of each species downstream the thinning LS, we integrate the concentrations within the $z$-dependent LS thickness given in Eq.~\ref{eq:thick}:
\begin{equation}
    f_{HDO}(t) = \frac{\int_{-d(t)/2}^{d(t)/2} f_{HDO}(x,t) dx}{A(t)},
    \label{eq:sim_HDO_conc}
\end{equation}
where $A(t)$ warrants proper normalization, i.e., $A(t)=\int_{-d(t)/2}^{d(t)/2}(f_{HDO}+f_{D_20}+f_{H_2O})dx$. The results for fitting Eq.~\ref{eq:sim_HDO_conc} to the experimental data as described in the main text are shown in Fig.~\ref{fig3}E.


\section*{Acknowledgments}
The authors appreciate discussions with Dr. Clemens Richter, Dr. Martin Thämer, Prof. Thomas Risse, and Dr. Kevin Wilson.

\paragraph*{Funding:}

N.S.M. acknowledges funding from the Deutsche Forschungsgemeinschaft (DFG, German Research Foundation) - Projektnummer 551280726. D. S., Ha. B., and N. H.
acknowledge financial support from the Cluster of Excellence ’CUI: Advanced Imaging of Matter’ of the Deutsche Forschungsgemeinschaft (DFG) – EXC 2056 – project ID 390715994. B.W. acknowledges funding from the European Research Council (ERC) under the European Union’s Horizon 2020 research and innovation programme (grant agreement No. 883759, AQUACHIRAL).

\paragraph*{Author contributions:}
TB and AP wrote the manuscript with feedback of all authors. NSM, GC, TB, HH, HaB, DS, NH, and AP performed the experiments. NSM, AP and TB analyzed the data. SG, MDP, and WS operated the free-electron laser. BW, MW, HeB, NH, GM, and AP supervised the project.

\paragraph*{Competing interests:}
There are no competing interests to declare.

\clearpage 
\printbibliography

\newpage


\renewcommand{\thefigure}{S\arabic{figure}}
\renewcommand{\thetable}{S\arabic{table}}
\renewcommand{\theequation}{S\arabic{equation}}
\renewcommand{\thepage}{S\arabic{page}}
\renewcommand{\thesection}{S\arabic{section}}
\setcounter{figure}{0}
\setcounter{table}{0}
\setcounter{equation}{0}
\setcounter{section}{0}
\setcounter{page}{1} 
\setcounter{secnumdepth}{1}


\begin{center}
\section*{Supplementary Materials for\\ Non-equilibrium state during proton-deuteron exchange at a liquid-liquid interface}


\author{
	Tillmann~Buttersack$^{\ast}$,
	Niclas~Sven~Mueller,
	Giulia~Carini,
    Henrik~Haak,
    Hanna~Bordyuh,
    Dipali~Singh,
    Sandy~Gewinner,
    Marco~De~Pas,
    Wieland~Schöllkopf,
    Martin~Wolf,
    Hendrik~Bluhm,
    Nils~Huse,
    Bernd~Winter,
    Gerard~Meijer,
    Alexander~Paarmann$^{\ast}$\and
    
	\small$^\ast$Corresponding author. Email: buttersack@fhi.mpg.de or alexander.paarman@fhi.mpg.de\and
 }

\end{center}

\newpage

\begin{figure}
    \centering
    \includegraphics[width=1\linewidth]{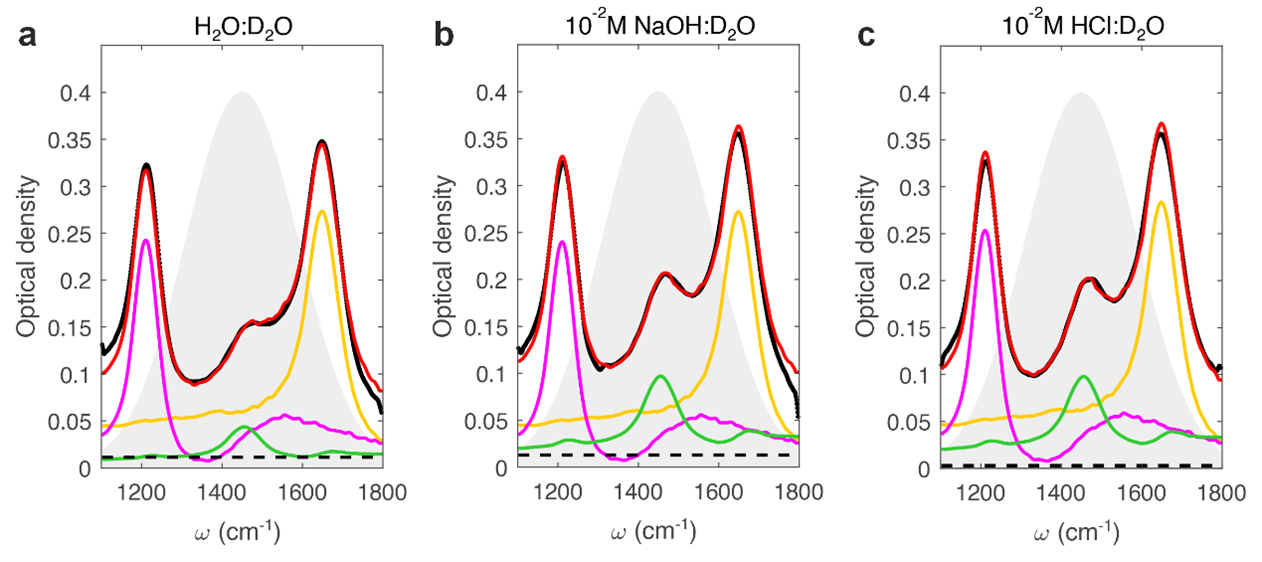}
    \caption{Fit of spectra from (a) \ce{H2O}:\ce{D2O}, (b) 10$^{-2}$\,M NaOH:\ce{D2O}, and (c) 10$^{-2}$\,M HCl:\ce{D2O} flat jets. Black curve shows measured data and red curve shows fit using a linear combination of reference spectra from \ce{H2O}:\ce{H2O} (orange), \ce{D2O}:\ce{D2O} (magenta), a 50:50 mixture of HDO (green), and a constant background (black dashed). The fit is weighted by a Gaussian centered at 1450\,cm$^{-1}$ (gray area). All spectra are obtained from an average of the lower half of the flat jet.}
    \label{fig:FigS2}
\end{figure}

\clearpage
\newpage

\section{Formation of HDO}
\label{sec:reactions}
The formation of HDO by mixing of H$_2$O and D$_2$O involves always hydronium or hydroxide ions. Concerted proton exchange is only debated in confined water. Since only one proton is transferred per step the reaction between H$_2$O  and D$_2$O leads to the formation of OH$^-$ and HD$_2$O$^+$ or OD$^-$ and H$_2$DO$^+$, respectively. Possible reactions leading to the formation of HDO are:

\begin{enumerate}
    \item \ce{OH^- + D2O -> HDO + OD^-}
    \item \ce{OD^- + H2O -> HDO + OH^-}
    \item \ce{H2DO+ + D_2O -> 2/3 HDO + 1/3 H2O + 2/3 HD_2O^+ + 1/3 D3O^+}
    \item \ce{HD2O+ + D_2O -> 2/3 HDO + 1/3 D2O + 2/3 D_3O^+ + 1/3 HD2O^+}
    \item \ce{H2DO+ + H_2O -> 2/3 HDO + 1/3 H2O + 1/3 H_2DO^+ + 2/3 H3O^+}
    \item \ce{HD2O+ + H_2O -> 2/3 HDO + 1/3 D2O + 2/3 H_2DO^+ + 1/3 H3O^+}
    \item \ce{OH^- + D3O+ -> HDO + D2O}
    \item \ce{OH^- + HD2O+ -> 4/3 HDO + 1/3 H_2O + 1/3 D_2O}
    \item \ce{OH^- + H2DO+ -> HDO + H_2O}
    \item \ce{OD^- + H3O+ -> HDO + H2O}
    \item \ce{OD^- + HD2O+ -> HDO + D_2O}
    \item \ce{OD^- + H2DO+ -> 4/3 HDO + 1/3 H_2O + 1/3 D_2O}
\end{enumerate}
For statistical reasons the reactions, which require hydroxide and hydronium ion to react (7-12) have a negligible impact on the HDO formation rate. 

\section{Fitting of the reaction-diffusion system to the data}

\begin{figure}[hbt!]
    \centering
    \includegraphics[width=1\linewidth]{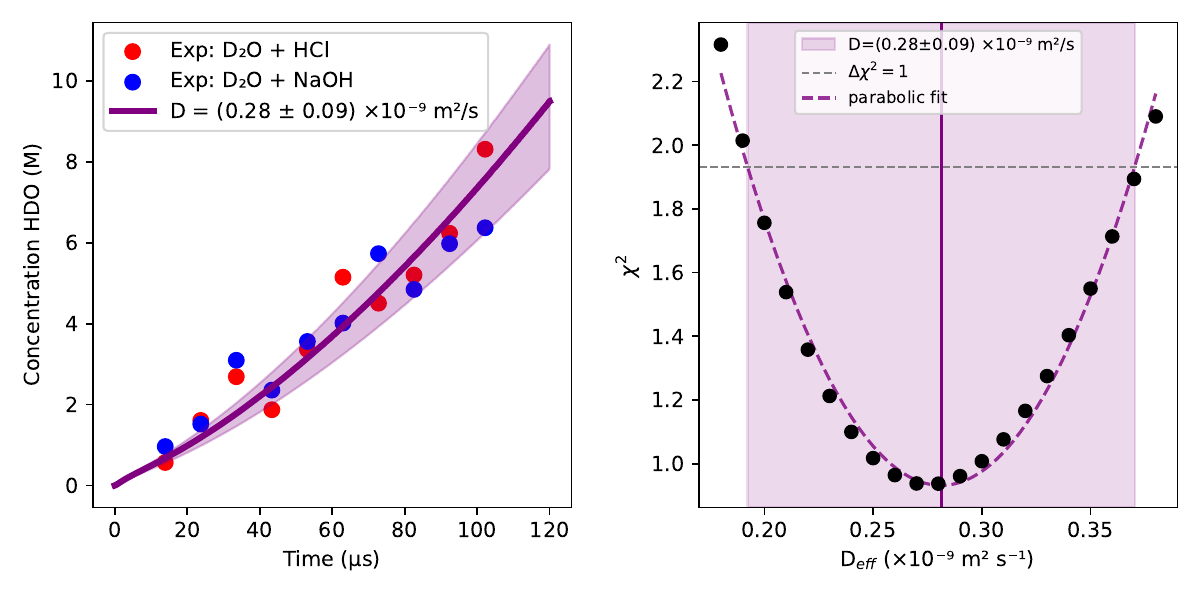}
    \caption{Fitting of diffusion constant (fast model): In the system (\ce{D2O} + 0.01\,M HCl or NaOH) the rate constant is too high to be measured with this system and it is assumed the local equilibrium is always reached. Here, only the diffusion coefficient is fitted. Right: standard deviation of data with different diffusion coefficients from experimental values.}
    \label{fig:SI_fit_D}
\end{figure}

First, we fitted data assuming a fast reaction ($k=10^9$  s$^{-1}$) and determined the diffusion constant $D_{\mathrm{eff}}$, meaning in each time step the local equilibrium is reached (see Fig.\,3F,H).
For this we calculated the HDO concentration by varying the effective diffusion $D_{\mathrm{}{eff}}$ and determined the reduced chi-squared of the diffusion only fit parameter $D_{\mathrm{eff}}$ ,(see figure\,\ref{fig:SI_fit_D}b)). 

\begin{equation}
    \chi^2 = \frac{1}{N} \cdot \sum_{i=1}^N \frac{\left(c_{HDO,exp} - c_{HDO,model}\right)^2}{\sigma_{D_{\mathrm{eff}}}} 
\end{equation}

The standard deviation $\sigma_{D_{\mathrm{eff}}}$ is determined by $\frac{1}{\sqrt{a}}$, where $a$ is the quadratic fit parameter (dashed line). The purple line in figure\,\ref{fig:SI_fit_D}) marks the fit ($D_{\mathrm{eff}} = (2.8\pm0.9) \cdot 10^{-10}$\,m$^{2}$ s$^{-1}$), while the light purple area indicates the confidence range of the effective diffusion coefficient.

Diaphragm-cell measurements report values for the self diffusion of HDO in \ce{H2O} and \ce{D2O} of 2.1 and 1.7$\cdot\,10^{-9}$\,m$^{2}$ s$^{-1}$, respectively.\cite{Mills1973} Deviations from these value can be ascribed to the uncertainty of the jet velocity and exact knowledge of the temperature, since the flat jet (FJ) was run in a open system without humidity control, such that a little evaporative cooling can not be excluded. Most important, the effective diffusion constant could be different from those literature values, as the FJ system is not static but rather fast flowing in one direction, as well as non-trivial mass transport within the LS and into the rim.

\begin{figure}[htp!]
    \centering
    \includegraphics[width=1\linewidth]{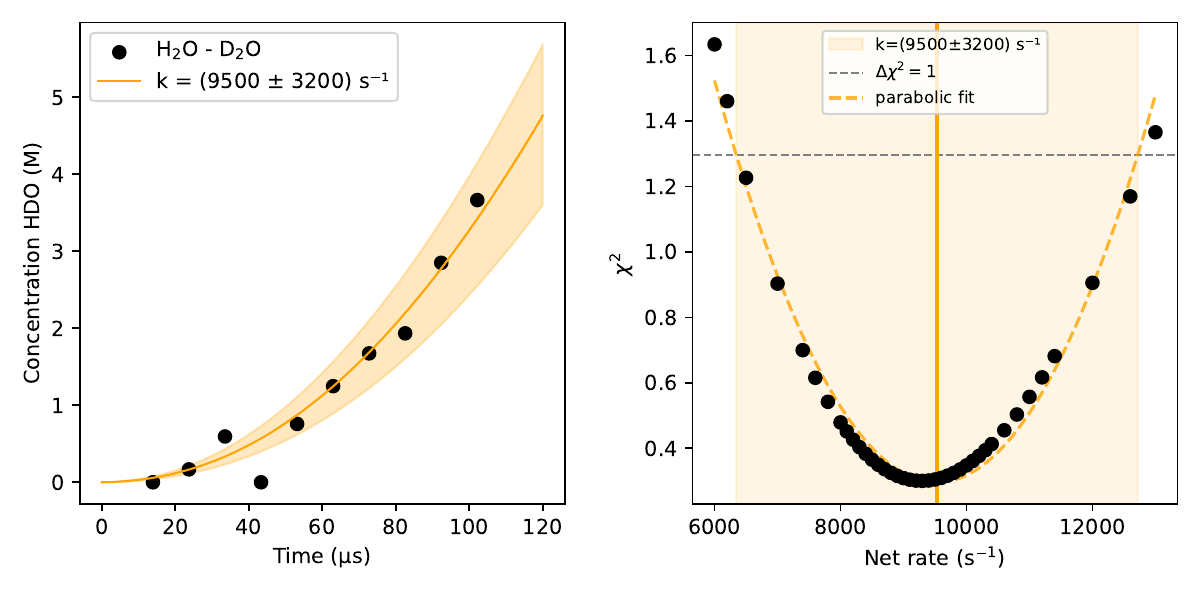}
    \caption{Fitting of the rate constant $k$ (slow model): For the case of pure \ce{H2O} and \ce{D2O} the diffusion constant is assumed to be as determined in the fast model. Right: standard deviation of data with different diffusion coefficients from experimental values.}
    \label{fig:SI_fit_teq}
\end{figure}

\begin{figure}[htp!]
    \centering
    \includegraphics[width=0.75\linewidth]{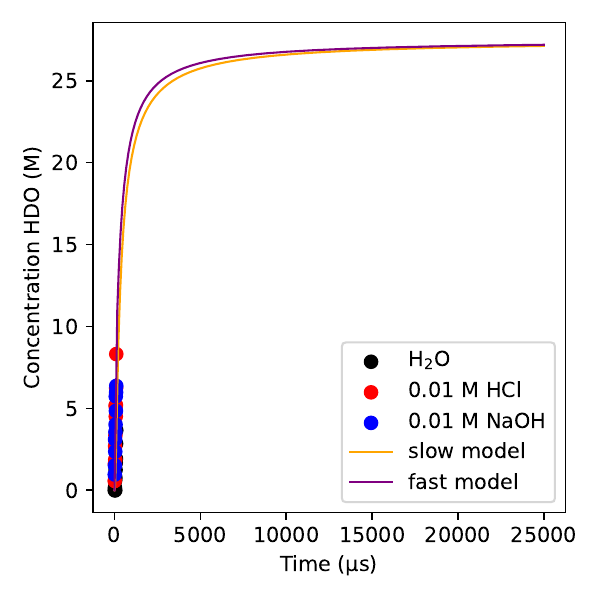}
    \caption{Modeled HDO concentration (as in Fig.\,3) over extended period of time. After a few ms in both cases the equilibrium concentration for 1:1 \ce{H2O}:\ce{D2O} mixture is reached.}
    \label{fig:short_vs_long}
\end{figure}

Second, we used a slow model, in which we assume the same diffusion constants of \ce{H2O}, \ce{D2O}, and HDO as fitted before with the fast model. Now, we varied the reaction rate and determined  analogously the reaction rate constant $k = 9500\pm3200$ s$^{-1}$ (Fig.\,\ref{fig:SI_fit_teq}).

The fitted curves (thicker purple and orange lines in Figs.\,\ref{fig:SI_fit_D} and \ref{fig:SI_fit_teq}) are displayed in the main part of the manuscript in Fig.\,3 as slow and fast model. In that figure, the time axis ends at 0.125\,ms. We repeated the calculations as described above over an extend time(25\,ms) showing that the concentration of HDO reaches as expected the equilibrium global concentration of HDO in both cases within a few microseconds. After that time the formation of HDO is diffusion limited independent of the hydronium and hydroxide concentrations. 

\newpage
\section{Further experiments}
\begin{figure}[hbt!]
    \centering
    \includegraphics[width=0.9\linewidth]{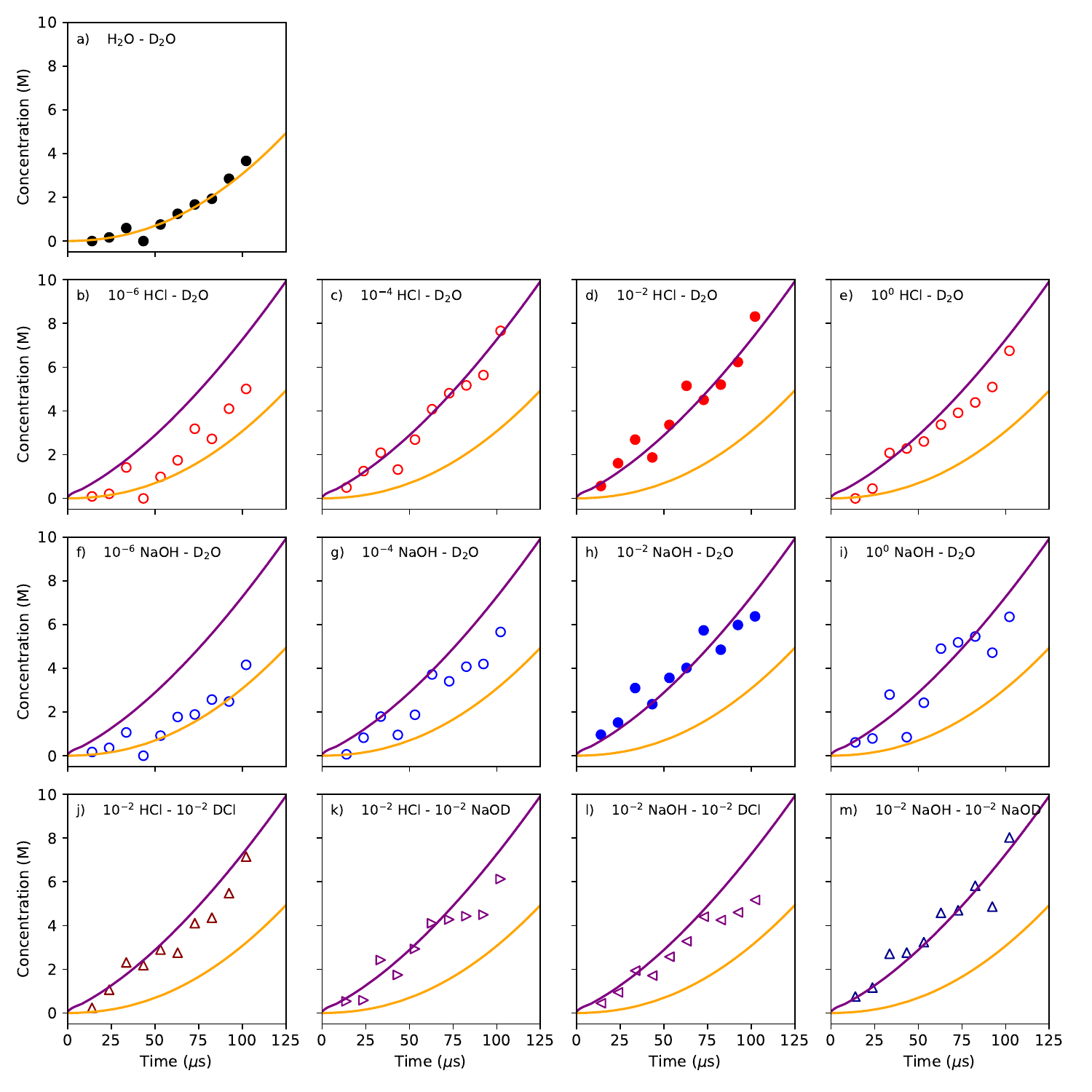}
    \caption{Concentration of HDO for different concentrations of acid and base (in mol/L). The data plotted in a, d, and h are discussed in the main manuscript are here only replotted for completeness. The purple and orange curve represent, as described in the main manuscript, the slow and fast models.}
    \label{fig:16plots}
\end{figure}

We have performed experiments also with other acid and base concentrations (1\,$\mu$mol, 0.1\,mmol, 10\,mmol, 1\,M) as shown in Fig.\,3A. The determined HDO concentrations are summarized in Fig.\,\ref{fig:16plots}. The fast model fits the majority of the datasets well.
Only in the cases b) and f), where the concentration of acid or base is as low as 0.001 mmol, the observed HDO concentration can be better modeled with the slow model (orange curve).

\clearpage 

\end{document}